\documentclass[runningheads]{llncs}
\usepackage[T1]{fontenc}

\usepackage{cite}
\usepackage{amsmath,amssymb,amsfonts}
\usepackage{graphicx}
\usepackage{textcomp}
\usepackage{xcolor}
\usepackage{subcaption}
\usepackage{multirow} 

\begin{document}

\def\deepspeech{${\tt DeepSpeech}$}
\def\DS{\deepspeech}
\def\t{\tau}
\def\alphabetset{{\cal A}}
\def\myspace{\square}
\def\myquote{\prime}
\def\mychange#1{{{#1}}}
\def\unk{\diamond}
\def\nmunk{\diamond}
\def\spoken#1{\vbox{/{\small {\tt #1}/}}}
\def\aps{{\sc aps}}
\def\wps{{\sc wps}}
\def\apw{{\sc awl}}
\def\active{${\tt Active}$}
\def\inactive{${\tt InActive}$}
\def\DSm{{\sc ds}}
\def\tim960{{25}}
\def\airdata{{\sc air-db}}
\def\allsstar{{\sc allsstar-db}}

\title{A Novel Scheme to classify Read and Spontaneous Speech}
 
\author{Sunil Kumar Kopparapu\inst{1}\orcidID{0000-0002-0502-527X}}
\authorrunning{Sunil Kumar Kopparapu}
\institute{TCS Research, India \\
\email{sunilkumar.kopparapu@tcs.com}\\
\url{http://www.tcs.com} 
}

\maketitle

\begin{abstract}
The COVID-19 pandemic has led to an increased use of remote telephonic interviews, making it important to distinguish between scripted and spontaneous speech in audio recordings. In this paper, we propose a novel scheme for identifying read and spontaneous speech. Our approach uses a pre-trained \DS\ audio-to-alphabet recognition engine to generate a sequence of alphabets from the audio. From these alphabets, we derive features that allow us to discriminate between read and spontaneous speech. Our experimental results show that even a small set of self-explanatory features can effectively classify the two types of speech very effectively.
\end{abstract}


\section{Introduction}
\label{sec:introduction}
The ability to automatically distinguish read speech\footnote{also called "prepared speech" or "scripted speech"} from spontaneous speech has several real world application. The pandemic introduced constraint on physical travels while there was no such constraint in terms of office work, especially because of the new paradigm of {\em work from  home}. As a result, people saw an opportunity to work for a organization that was hitherto not on their radar because of physical distance. The need to travel to work constraint removed, all work places were an opportunity  as a result there was a large movement of people across organizations. 
% This was both, an opportunity to organizations to get the best people on-board and it was also an opportunity for people to scout for better work. 
The shift to remote work during the pandemic created opportunities for both organizations to hire top talent and for individuals to explore new job prospects.
Any movement into an organization is preceded by an {\em interview} and in the remote work scenario these were in the form of 
audio or telephone based interviews. Given the large volume of people who were crisscrossing, several organization used semi-automated methods to conduct interviews, especially to filter out the initial applicants. One of the critical aspect that required monitoring was to determine {\em if the candidate was responding to the question spontaneously or was she reading from a prepared or scripted text}. The need for an automatic identification of  the candidate speech during interview as read speech or spontaneous speech became necessary. In another use case, the ability to distinguish read-speech and spontaneous-speech can have applications in forensics to distinguish {\em "asked to read"} statement (or confession) from spontaneous statement of a person being investigated. This can possibly be useful to determine if the statement given by the person was given on {\em own accord} or was forced to give the statement.

There have been several approaches adopted by researcher in the past which dwell into classification of read and spontaneous speech. Most of these approaches have used deep and intricate analysis of the audio signal %and $|$ 
or language or both 
to distinguish read and spontaneous speech. More recently, pivoting on fluency in L2 language,  \cite{eren2021fluency} studies the essential statistical differences, based on data collected, in pauses between read and spontaneous speech, for  Turkish, Swahili, Hausa and Arabic speakers of English.  In \cite{DELLWO201513}, the authors describe method to recognize read and spontaneous in Zurich German (a specific dialect spoken in Switzerland) language. The authors in \cite{batliner1995can} discuss the possibility of differentiation between read and spontaneous speech by just looking at the intonation or prosody. Read and spontaneous speech classification based on variance of GMM supervectors has been studied in \cite{asami2014read}. From a speaker role characterization perspective, in \cite{Dufour2014CharacterizingAD} the authors use acoustic and linguistic features derived from an automatic speech recognition system to characterize and detect spontaneous speech. They demonstrate their approach on three classes of spontaneity labelled French Broadcast News. 

Two unrelated works reported in literature three decades apart influence the novel approach proposed in this paper. The first one is an  early work on understanding spontaneous speech \cite{Ward_understandingspontaneous}. It captures the essential differences between read and spontaneous speech while trying to reason out why systems, like automatic speech to text recognition, designed to work for read speech often fail to perform well on spontaneous speech. %Some of the idiosyncrasies associated with spontaneous speech have been highlighted. 
They equate read speech to written text and spontaneous speech to spoken speech and highlight some of the idiosyncrasies associated with spontaneous speech.  
% To quote \cite{Ward_understandingspontaneous}, 
% {\em  ``People use language differently when they speak than when they write. Spoken language
% contains many interjections, filled pauses, etc. Speakers often don't use well-formed sentences. They speak in
% phrases, have restarts, etc.''}. 
Though the authors intent was 
%to highlight that systems designed for read speech   
%Systems designed for written or read text will encounter serious difficulties processing
%such input. This paper 
to outline strategies for speech recognition
system trained for read speech to deal with spontaneous spoken speech, it captures some crucial differences in read and spoken speech which can be very helpful in building a classifier to distinguish read and spontaneous speech.
 Though not directly related to read and spontaneous speech, the second influence is the work reported in \cite{TRIPATHI2021101213} where 
 %. The authors in  \cite{TRIPATHI2021101213} 
 they exploit the pre-trained \deepspeech\ speech-to-alphabet recognition engine to estimate the intelligibility of dysarthric speech. 
 %The approach adopted in this paper is motivated by this work.
 This paper is influenced by the approach adopted in \cite{TRIPATHI2021101213} to identify the differences between read and spontaneous speech as mentioned in \cite{Ward_understandingspontaneous}. More recently, \cite{9806803} made use of the differences between spoken language text  and written language text, derived from spontaneous and read speech respectively, to build a language model that enhances the performance of a speech to text  engine.
 
 The main aim of this paper is to introduce a novel approach to identify features that are not only self explanatory but are also able to distinguish between read and spontaneous speech. To the best of our knowledge, there is no {\em known} system to distinguish read and spontaneous speech in literature. Please note that, for this reason, we are unable to %make no attempt to 
 compare the performance of the approach proposed in this paper with any prior art.
 %another system in literature, primarily because we did not come across any existing system. 
 The essential idea is to exploit the available deep pre-trained models to extract features, from speech, that can discriminate between read speech from spontaneous speech. The rest of the paper is organized as follow: In Section \ref{sec:approach}, we describe our approach through an example. In Section \ref{sec:experiments}, we present our experimental results and conclude in Section \ref{sec:conclusions}.

\section{Our Approach}
\label{sec:approach}
\begin{figure}[ht]
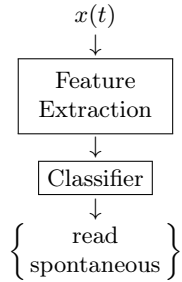

    \centering

  $x(t)$ 
  
  $\downarrow$
  
  \fbox{
  $\begin{array}{c} 
  \mbox{Feature}\\ 
  \mbox{Extraction}
  \end{array}$
  }
  
   $\downarrow$
  
  \fbox{\mbox{Classifier}}
  
  $\downarrow$
  
  $\left \{\begin{array}{c}
  \mbox{read} \\
  \mbox{spontaneous} \end{array} \right \}$
 \caption{A high-level read and spontaneous speech classification scheme.}
    \label{fig:classifier}
\end{figure}

The problem of read and spontaneous speech classification can be stated as \begin{quote} \em Given a recorded audio sample, spoken by a single person, $x(t)$, determine automatically if $x(t)$ was {read} or spoken {spontaneously}.
\end{quote}
While the approach is simple and straightforward as seen in %Our approach is captured 
in Fig. \ref{fig:classifier}, the novelty is in the feature extraction block that utilizes {\em unconventional}, yet explainable set of features,  that  aid distinguish { read} and {spontaneous} speech. Additionally, this %set of 
features are easily %explainable and are 
obtained using \DS\ a pre-trained speech-to-alphabet recognition engine \cite{MozillaDS}.

\subsection{Speech-to-Alphabet (\DS)}

Mozilla's \DS\ \cite{MozillaDS}  is an end-to-end deep learning model 
that converts
speech into alphabets based on the Connectionist Temporal Classification (CTC) loss function. The $6$ layer deep model is pre-trained on $1000$ hours of speech from
the Librispeech corpus \cite{panayotov2015librispeech}. All the $6$ layers, except
the $4^{th}$,  have feed-forward dense units; the $4^{th}$ layer itself has recurrent units.

A speech utterance $x(t)$ is segmented into $T$ frames, as is common in speech
processing, namely,  $x^{\t}({t})\ \ \forall \t \in \left[0, T-1\right]$. In \DS, each frame is of duration $\tim960$ msec.
Each frame $x^{\t}(t)$ is represented by $26$
%Mel-filter Cepstral Coefficients (
Mel Frequency Cepstral Coefficients \mychange{(MFCCs)}, denoted by $\vec{f}_\t$.
Subsequently, the complete speech utterance $x(t)$ can be represented as %a
%sequence of speech features, namely, 
$\{\vec{f}_\t\}_{\t=0}^{T-1}$. The input to
%the 
\DS\ is %the speech features from 
$9$ preceding and $9$ succeeding frames,
namely $\{ \vec{f}_{\t-9}, \cdots, \vec{f}_{\t+9}\}$.
The output of the \DS\ model is a probability distribution over an
alphabet set 
%$\alphabetset$ of a  particular language.
%In case of English language, 
$\alphabetset = (a, b, \cdots, z, \nmunk,  \myspace, \myquote)$ with
$|\alphabetset| = 29$. Note that there are three additional
outputs, namely, $\unk$, $\myspace$, and $\myquote$ corresponding to {\em unknown}, {\em space} and
an {\em apostrophe}, respectively in $\alphabetset$ in addition to the $26$ known English alphabets\footnote{a collection of letters $\{a, b, \cdots, z\}$ }.
The output at each frame, $\t$ is
\begin{equation}
c^*_{\t} = \max_{\forall  k \in \alphabetset} P\left(\left(c_{\t} = k\right)|\left \{\vec{f}_{\t-9}, \cdots, \vec{f}_{\t}, \cdots, \vec{f}_{\t+9}\right \}\right) %\;\;
\label{eq:mozds}
\end{equation}
where $c^*_{\t} \in \alphabetset$. It is important to note that a typical
speech recognition engine is assisted by a statistical language model (SLM or  LM for short), 
which helps in {\em masking} small acoustic mispronunciations. However, as seen in
(\ref{eq:mozds}), there is no role of LM.
% in the process converting s2a.
 This, as we will see
later, helps in our task of extracting features that can assist distinguish read and spontaneous speech. As we mentioned earlier, the use of \DS\ is motivated by its use for speech intelligibility estimation work reported in \cite{TRIPATHI2021101213}. 
Note that (a) \DS\ outputs an alphabet for every frame of $\tim960$ msec, so the longer the duration of the audio utterance, the more the number of output alphabets, (b) the output is always from the finite set $\alphabetset$ based on Equation (\ref{eq:mozds}).
Note that $\myspace$ can be treated as the {\em word separator} and we refer to $\unk$ %(also know as {\em unknown} 
token in \DS\ as an {\inactive} alphabet and anything other than that, namely, $\left \{{\cal A}\right \} - \unk$  as the {\active} alphabet.
    
\subsection{Feature Extraction}

An example the raw output of \DS\ to an utterance $x(t) = $
%\begin{quote}
\begin{equation*}
%x(t) =
%\mbox
{\spoken{Declaration of a variable is merely specifying the data}}
\end{equation*} 
%\end{quote}
is $\mbox{\DSm}(x(t)) =$
\begin{quote}

$\unk \unk \unk\unk\unk\unk\unk\unk\unk\unk\unk\unk\unk\unk\unk\unk\unk\unk\unk\unk\unk\unk\unk\unk\unk\unk d\unk e\unk\unk\unk\unk c\unk\unk a\unk\unk r\unk\unk\unk\unk i\unk\unk\unk\unk tiio\unk n\unk\unk\unk\unk\myspace
\unk\unk o\unk f\unk\unk\myspace
\unk\unk a\unk\unk\myspace
r\unk e\unk\unk l\unk\unk i\unk\unk\unk aa\unk\unk\unk b\unk le\unk\unk\unk\unk\unk\unk\unk\myspace
\unk\unk\unk\unk i\unk\unk ss\unk\unk\unk\unk\unk\myspace
\unk\unk m\unk\unk\unk e\unk\unk\unk r\unk e\unk\unk\unk\unk l\unk y\unk\unk s\unk\unk\unk\unk\myspace
\unk p\unk\unk e\unk c\unk\unk\unk\unk i\unk\unk\unk\unk f\unk\unk\unk\unk y\unk\unk iing\unk\unk\myspace
\unk thhat\unk\myspace
\unk\unk\unk\unk e\unk\unk\unk\unk\unk t\unk\unk a\unk\unk\unk\unk\unk\unk\unk\myspace
$
    
\end{quote}

\DS\ raw output of an audio signal $x(t)$ is a string of alphabets ($\in \alphabetset$). In this paper, we assume $\mbox{\DSm}(x(t))$ to represent the audio signal $x(t)$ and hence any signal processing required to extract features from the audio signal translates to simple {\em string} or text processing. As seen from $\mbox{\DSm}(x(t))$, we can easily extract several features using simple string processing scripts. For example, 
%a count 
the number of words in the spoken utterance can be identified by the number of occurrences of $\myspace$. We can count the total number of alphabets, the total number of \inactive\ and \active\ alphabets by processing the alphabet string. Additionally, the knowledge of the duration of the audio $x(t)$ means that we can compute velocity-like features, for example, alphabets per second (\aps) or words per second (\wps) etc or number of \inactive\ or \active\ alphabets per sec or number of active average word length (\apw) or alphabets per word and so on. 

We hypothesize that $\mbox{\DSm}(x(t))$, as a representation of speech $x(t)$, contains sufficient information that can help distinguish between read and spontaneous speech along the lines of \cite{Ward_understandingspontaneous}. This is motivated by the fact that given the {\em same information} to be articulated by a speaker, read speech is much faster compared to spontaneous speech, meaning the {\em duration} of the spontaneous speech is much longer than the read speech. 
%Considering the fact that a word has a predefined number of phonemes (or alphabets in the case of \DS) 
If we consider that spontaneous speech requires thinking time between words, between sentences \cite{Ward_understandingspontaneous} etc then the number of \inactive\ alphabets must be more in spontaneous speech compared to read speech. Namely, for the same sentence, 
%if we 
%which again translates to %the fact that 
the output of \DS\ should 
having more number of \inactive\ alphabets compared to read speech.
%for a given word or 
% in the utterance. It is to be noted that these features that we are speaking about are explainable.
\begin{figure}[ht]
    \centering
    \includegraphics[width=0.85\textwidth]{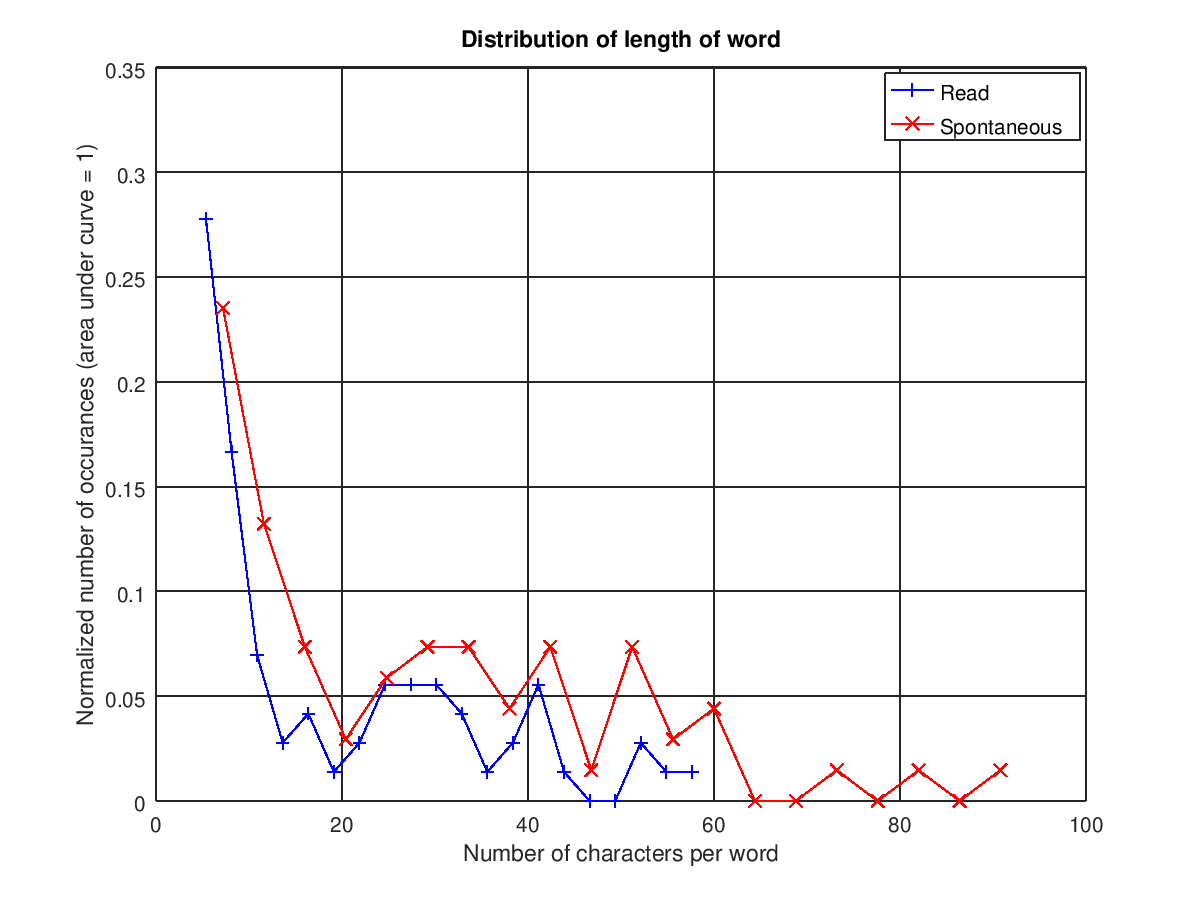}
    \caption{Word length (\# of alphabets per word) for read and spontaneous speech.}
    \label{fig:word_length}
\end{figure}

\subsection{Identifying Features}
In the highly data-driven machine learning era, we opted to look for simple, yet effective features that could help in our pursuit. We considered a short technical passage consisting of two sentences and $62$ words, which we picked from Wikipedia for our analysis 
% \begin{quote} \em
    % "Declaration of a variable is merely specifying the data type of a variable and the
% variable name. As a result of the declaration we merely tell the compiler to reserve the
% space for a variable in the memory according to the data type specified. Whereas a
% definition is an implementation of the declared variable where we tie up
% appropriate value to the declared variable so that linker will be able to link references
% to the appropriate entities."
% \end{quote}
and asked (a) the paragraph to be read as is (read speech) and (b) the paragraph to be held as a reference and spoken in their own words ($\equiv$ spontaneous). We recorded this on a laptop as a $16$ kHz, $16$ bit, mono in {\tt .wav} format. This read and spontaneous audio was processed by \DSm$()$ to produce a string of alphabets ($\in \alphabetset$). 
Fig. \ref{fig:word_length} shows a histogram plot of the number of alphabets in a word and their normalized frequency (area under the curve is $1$). It can be clearly observed that, (a) there are more words (with same number of alphabets\footnote{we use letter, character and alphabet interchangeably}) in spontaneously spoken passage compared to the read passage (the plot corresponding to spontaneous speech, in red is always above the read speech) and (b)
there are more lengthy words in spontaneous speech (the spontaneous speech plot spreads beyond the read speech blue curve), there are words of length $90$ alphabets in spontaneous speech compared to $< 60$ alphabets per word in read speech. This is in line with the observation that there are more \inactive\ alphabets in spontaneous speech.

\begin{figure}[ht]
    \centering
    \includegraphics[width=0.85\textwidth]{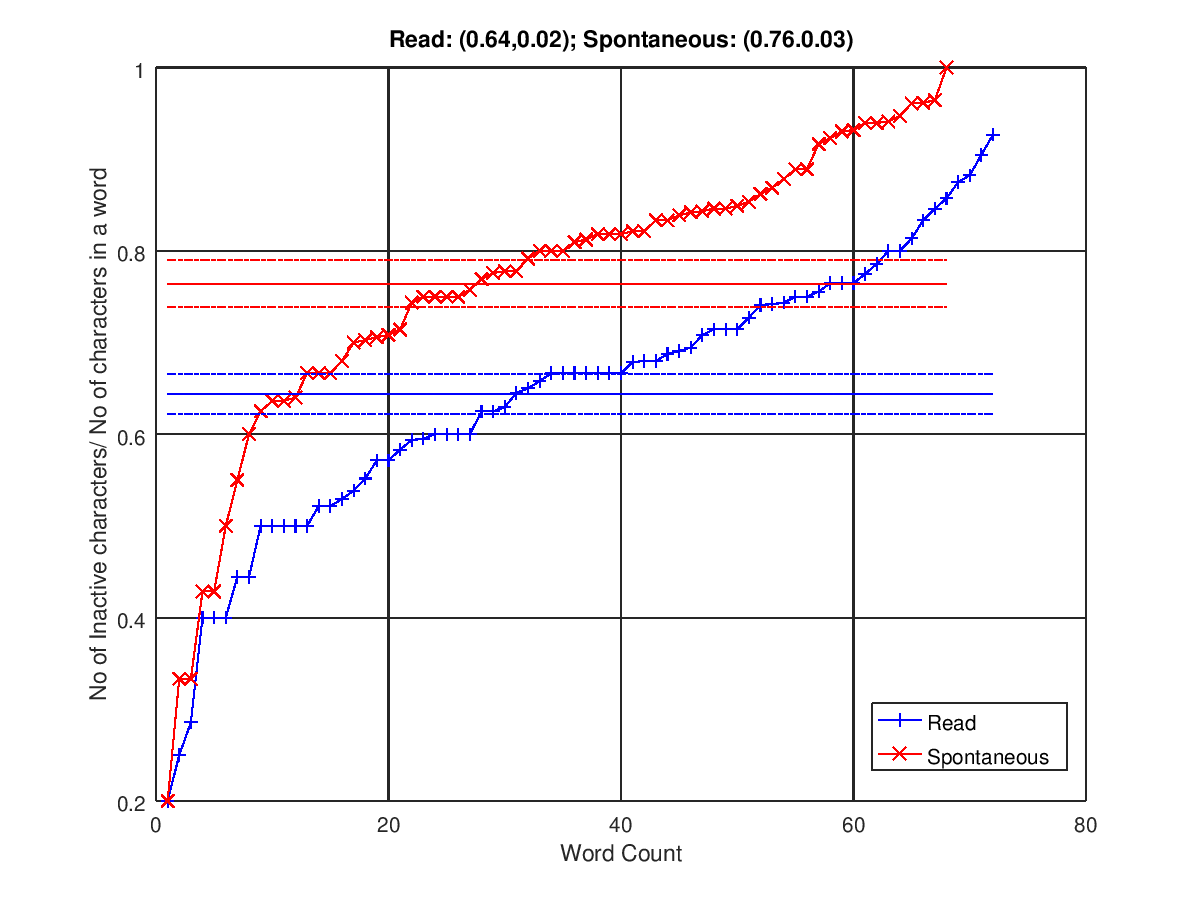}
    \caption{Ratio of \# \inactive\ alphabets to the \# of alphabets in a word (arranged in the increasing order of ratio).}
    \label{fig:inactive_alphabets}
\end{figure}

%Also, 
We extracted a set of $5$ meaningful features %\footnote{, since they are } 
as mentioned in Table \ref{tab:read_vs_spontaneous} for both the read and spontaneous speech. Note that these measured features are self explanatory and so we do not describe them in detail. Clearly, there are $3$ features (the duration (a), the number of alphabets (c), and the number of \active\ alphabets (d)) that show promise to discriminate %distinguish the 
read and the spontaneous speech. %; for example, . 
\begin{table}[ht]
    \centering
%  \resizebox{0.5\textwidth}{!}
{  
 \begin{tabular}{c|l|c|c}  \hline
 \multicolumn{4}{c}{Measured Values}\\ \hline
SNo &    What  &  Spontaneous & Read\\ \hline
(a)&   {Duration} (sec) & {{47.62}} & {{29.67}}  \\
(b) &    Number of Words (\#)  &69     & 72\\
(c) & {Number of Alphabets}  (\#) & 2382  &  1484\\ 
(d) & {Number of \active\ alphabets} (\#)  &  1915 & 951 \\
(e) & Number of \inactive\ alphabets (\#)  & 364 & 413  \\ \hline\hline
 \multicolumn{4}{c}{Derived Features}\\ \hline
 Ratio &    What  &  Spontaneous & Read\\ \hline
${(c)}/{(b)}$ & Av word len (alphabets/word; \apw) &34.52       & 20.61    \\     
${(c)}/{(a)}$ & Speaking Rate (alphabets/sec; \aps)    &    50.02  &      50.02  \\
${(b)}/{(a)}$ & {{Word Rate (\wps)}} $[f_3]$     & {{1.45}}   &    {{ 2.43}}  \\
${(e)}/{(a)}$& {{\inactive\ \aps}} $[f_2]$& {{7.63 }}& {{13.92}}\\ 
${(d)}/{(b)}$& {{\active\ \apw}} $[f_1]$& {{27.75}} & {{13.21}} \\ \hline\hline
 \end{tabular}
 }
    \caption{Measured features from read and spontaneous speech for the same paragraph. \# denotes is the count, an integer.}
    \label{tab:read_vs_spontaneous}
\end{table}
Based on %our %interpretation of \cite{Ward_understandingspontaneous} 
%understanding of 
the differences between read and spontaneous speech mentioned in \cite{Ward_understandingspontaneous} we %Additionally, we extract some 
derive (see %derived features %as shown in 
Table \ref{tab:read_vs_spontaneous} Derived Features) features like average word length (\apw), speaking rate, word rate, \inactive\ \aps\ and \active\ \apw, from the values directly measured from $\mbox{\DSm}(x(t))$. 
It can be observed that, while {\active\ %alphabets per word
average word length} (\active\ \apw) and  {{\inactive\ alphabets per sec}} (\inactive\ \aps) features show promise to be able to discriminate read and spontaneous speech, the speaking rate in terms of alphabets per sec (\aps) is a feature that does not allow us to discriminate between read and spontaneous speech, this is to be expected because as we mentioned earlier, the total number of alphabets output by \DSm$()$ is proportional to the duration of the utterance\footnote{one alphabet for every $25$ msec}. Clearly, the \active\ and \inactive\ alphabets play an important %major 
role in discriminating read and spontaneous speech. As one would expect, there are a large number of %{\em think} pauses (resulting in more 
$\unk$ (can be associated with pauses) in spontaneous speech compared to read speech. 
Fig. \ref{fig:inactive_alphabets} shows the plot of the ratio of number of \inactive\ alphabets to the number of alphabets in a word (arranged in the increasing order). It can be observed that spontaneous speech has more \inactive\ alphabets per word compared to the read speech. Note that the curve corresponding to spontaneous speech, in red, is always higher than the read speech (blue curve). This is expected, considering that there is a sizable amount of pause time in spontaneous speech, unlike read speech. We can further observe that the means %average 
value of the ratio (number of \inactive\ alphabets to the number of alphabets) is higher for %computed for an utterance showed the 
spontaneous speech %to have a value of 
($0.76$) compared to read speech ($0.64$) as seen in % for read speech (see 
Fig. \ref{fig:inactive_alphabets}. %).

\subsection{Proposed Classifier}

%In summary, 
As observed in the previous section, there exist %both measured and derived 
features extracted from \DS\ that are able to discriminate read and spontaneous speech. 
However, the measured features (Table \ref{tab:read_vs_spontaneous} (a), (c), (d)) though able to discriminate read and spontaneous speech  are not useful because it requires %one is not 
{\em a priori} knowledge of %aware of 
the passage or information spoken by the speaker. On the other hand, 
there are a set of derived features, which are ratios and hence independent of the spoken passage. %derived %from the string of  alphabets output by \DS\ speech to alphabet engine 
%which are , 
As seen in Table \ref{tab:read_vs_spontaneous} some of these features are able to strongly discriminate read and spontaneous speech. The three derived features that show promise to discriminate read and spontaneous speech are 
\begin{enumerate}
    \item{[$f_1$]} \active\ \apw\ 
    
    (\active\ alphabets per word is higher for spontaneous speech)
    \item{[$f_2$]} \inactive\ \aps\ 
    
    (\inactive\ alphabets per sec is lower for spontaneous speech)
    \item{[$f_3$]} \wps\ 
    
    (Word Rate or Words per sec is lower for spontaneous speech)
\end{enumerate}
Note that these features are independent of the duration of the audio utterance and they do not depend on  {\em what was spoken} and entirely rely on {\em how the utterance was spoken}. This is important because any feature based on {\em what was spoken} would have a direct dependency  on the performance accuracy of the speech-to-alphabet engine, in our case \DS. 
In that sense our approach does not depend explicitly on the performance of the \DS\ and does not depend on the linguistic content of the spoken passage. % or information spoken.
The process of classifying a given utterance $u(t)$ is simple\footnote{there is no need to train a conventional classifier}. We extract the %three %derived 
features $f_1, f_2, f_3$ from the $\mbox{\DSm}(x(t))$ for a given spoken passage $x(t)$ and compute a read score ${\cal R}$ using (\ref{eq:classification}). We use (\ref{eq:threshold}) to determine if $x(t)$ is read speech or spontaneous speech.
\begin{equation}
   {\cal R} = \frac{1}{1+exp^{-\lambda_1(f_1-\tau_1)}} +
    \frac{1}{1+exp^{\lambda_2(f_2-\tau_2)}} + 
    \frac{1}{1+exp^{-\lambda_3(f_3-\tau_3)}}
    \label{eq:classification}
\end{equation}

%Once we compute ${\cal R}$, we 
\begin{eqnarray}
x(t) & =& \mbox{Read Speech}\;\; \mbox{if}\;\; {\cal R} \ge \tau_{{\cal R}}  \nonumber \\
 &=& \mbox{Spontaneous Speech}\;\; \mbox{if}\;\; {\cal R} < \tau_{{\cal R}}  
 \label{eq:threshold}
\end{eqnarray}
%
%In (\ref{eq:classification}) 
We empirically chose $\lambda_{1,2,3}=1$, 
$\tau_1 = 6$, 
$\tau_2 = 10$, and 
$\tau_3 = 1.75$ %are chosen 
based on observations made in Table \ref{tab:read_vs_spontaneous}. %and set the value of 
And 
$\tau_{{\cal R}}=1.75$, 
%the mid value  of
which is in the range
%the range of %based on the observation  
%that 
${\cal R} \in [0,3]$. 
%Note that ${\cal R}$ can take a value between $0$ and $3$.
%and $\tau_{{\cal R}}=2$ is chosen so that at least two of the three features indicate that the utterance is read speech. 

%If ${\cal R}$ to determine if $x(t)$ is read or spontaneous.

%{\em if and only if} the {\em same information} is uttered. This is impractical in reality because the speaker could have uttered. 

\section{Experimental Validation}
\label{sec:experiments}
\begin{figure}[htb!]
    \centering
    \begin{subfigure}{0.49\textwidth}
    \includegraphics[width=\textwidth]{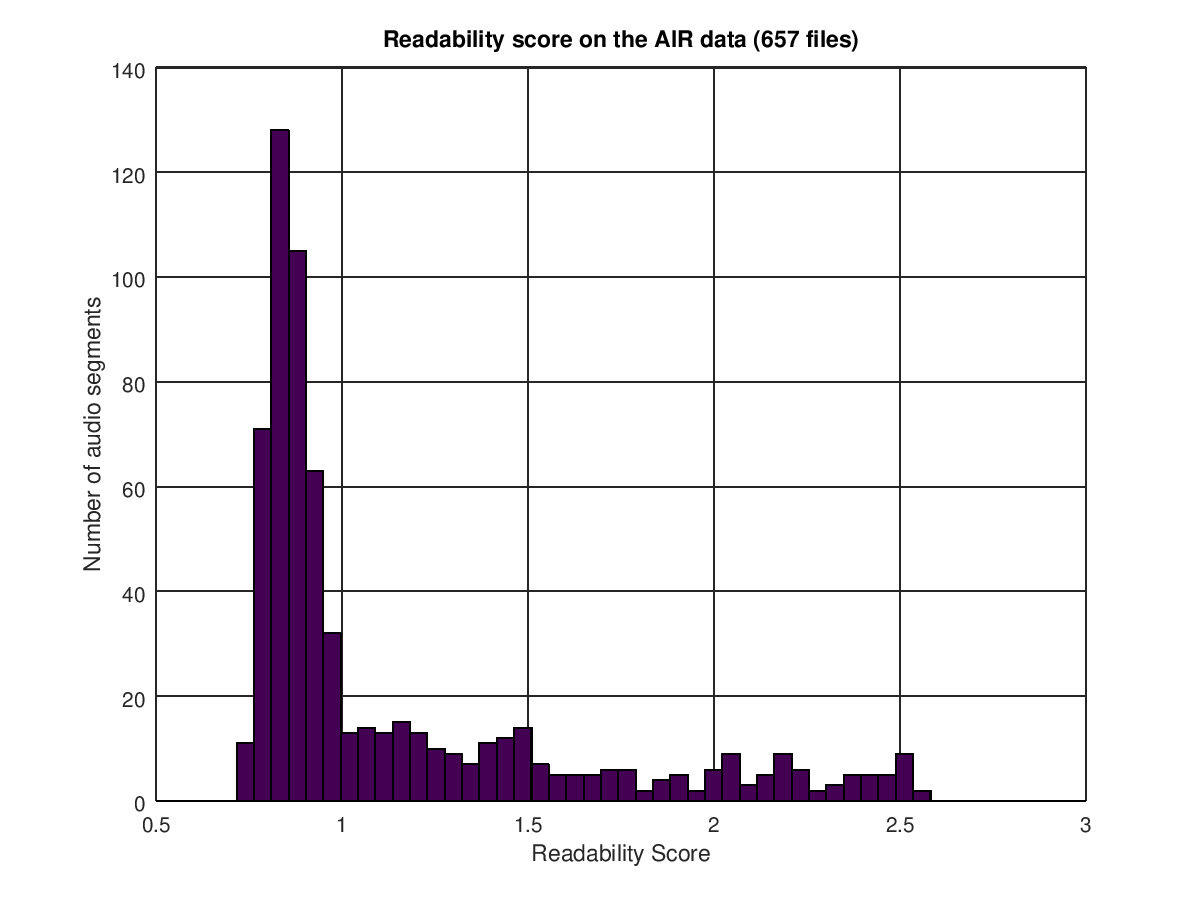}
    \caption{Histogram of ${\cal R}$ score.}%distribution on $657$ audio segments.}
    \label{fig:air_distribution}
\end{subfigure}
\hfill
\begin{subfigure}{0.49\textwidth}
    \includegraphics[width=\textwidth]{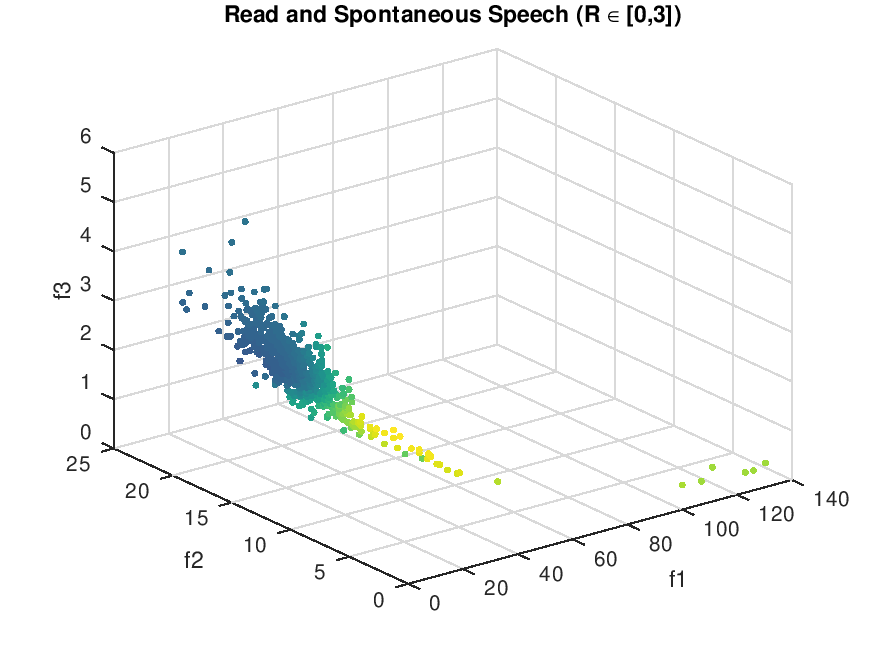}
    \caption{${\cal R}$ as a function of $f_1, f_2, f_3$.}
    \label{fig:scatter} 
\end{subfigure}
\caption{Readability score (${\cal R}$) for $657$ audio segments ($> 2$ sec) from \airdata.}
\label{fig:airdata}
\end{figure}

The selection of the features to discriminate between spontaneous and read speech is based on an intuitive understanding of the difference between read and spontaneous speech as mentioned in \cite{Ward_understandingspontaneous} and verified through observation of actual audio data (Table \ref{tab:read_vs_spontaneous}). %and analysis 
We collected audio data ($150$ minutes; spread over $7$ different programs) broadcast by All India Radio \cite{AIR} called %\footnote{https://newsonair.gov.in/} 
\airdata\ which is available at \cite{air-db}. 
This audio data is the recording between a host and a guest and consists of both spontaneous speech (guest) and read speech (host). We used a pre-trained speaker diarization model \cite{spkdiarization,Bredin2020}
%\footnote{pyannote/speaker-diarization@2022.07} 
to segment the audio, which resulted in $1028$ audio segments. We discarded all audio segments below $2$ sec so that there was sizable amount of spoken information in any given audio segment; this resulted in a total of $657$ audio segments. All experimental results are reported on this $657$ audio segments (see Fig. \ref{fig:airdata}).
% We are in the process of scouting for a %n available 
% data-set specifically build for distinguishing read and spontaneous speech to extensively test our simple, yet {\em novel set of explainable features} to classify speech into read or spontaneous speech.
% \begin{figure}
%     \centering
%     \includegraphics[width=0.85\textwidth]{air_distrubution.png}
%     \caption{}
%     \label{}
% \end{figure}

\begin{figure}
    \includegraphics[width=0.85\textwidth]{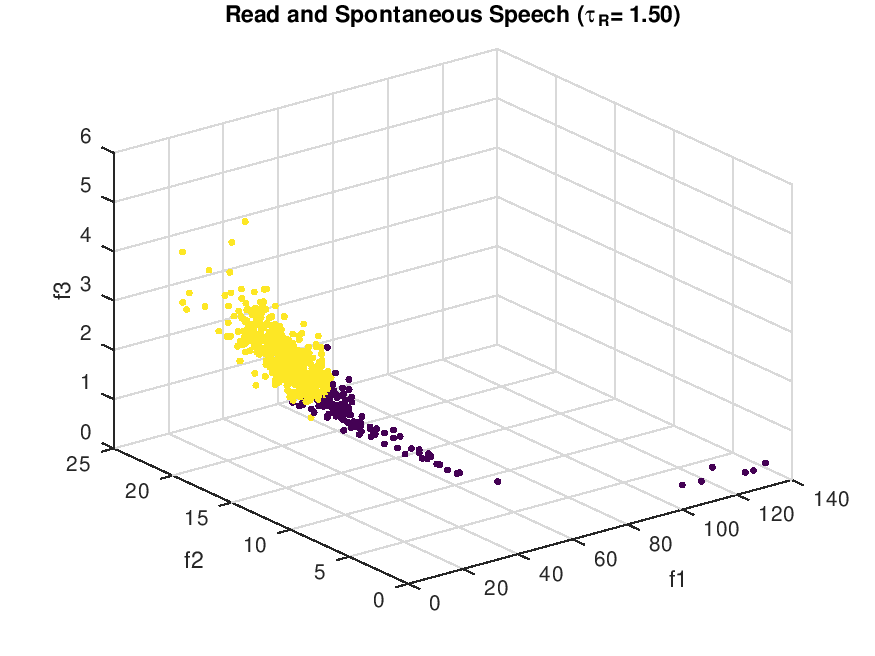} \hfill
    \caption{$657$ audio segments from \airdata\ classified as read speech (violet) and spontaneous speech (yellow).}
    \label{fig:classes}
\end{figure}

 For each of these $657$ audio segments, $f_1, f_2, f_3$ were computed and then using (\ref{eq:classification}) ${\cal R}$ was computed. Fig. \ref{fig:air_distribution} shows the distribution of the readability score ${\cal R}$ of the audio segments. Clearly a large number of  audio segments ($535$) were classified as spontaneous speech compared to $122$, which was classified as read. Figure \ref{fig:scatter} shows the scatter plot of ${\cal R}$ for the $657$ audio segments as a function of $f_1, f_2, f_3$. The colour of the scatter plot represents the value of ${\cal R}$. Figure \ref{fig:classes} shows the classification of segmented audio into read speech (violet; ${\cal R} \ge \tau_{\cal R}$) %=1.5$) 
 and spontaneous speech (yellow; ${\cal R} < \tau_{\cal R}$). %=1.5$

 We choose $\delta = 0.05$ and  selectively listen to some of the audio segments (${\cal R} > (\tau_{\cal R}+\delta)$ %=1.55$ 
 and ${\cal R} < (\tau_{\cal R}-\delta)$) % = 1.45$) 
 and found that almost all of the audio segments classified as spontaneous belong to the guest speaker (which is expected), however, several instances of host speech was also classified as spontaneous. 
 %On careful listening, it could be observed that the host (one would expect them to read from written text) also sounded spontaneous. 
 We hypothesize, that radio hosts are trained to speak even written text to give a feeling of spontaneity to the listener. We then looked at the $23$ audio segments which had ${\cal R}$ in the range $[\tau_{\cal R}-\delta, \tau_{\cal R}+\delta]$ and hence in the neighbourhood of $\tau_{\cal R}$ %= 1.5$ 
 which is more prone to classification errors. We observed that there were $12$ and $11$ read speech and spontaneous speech segments respectively. Of the $12$ audio segments classified as read speech, $4$ audio segments were actually spontaneous while of the $11$ audio segments classified as spontaneous speech, $3$ audio segments were actually read speech (see Table \ref{tab:delta_analysis}). It should be noted that, in the neighbourhood of the $\tau_{\cal R}$, where the confusion is expected to be very high, the proposed classifier is able to correctly classify with an accuracy of $\approx 70\%$ ($16$ of the $23$ audio segments correctly classified).

 \begin{table}[htb!]
     \centering
     \begin{tabular}{|c|c|c|} \hline
    ${\cal R} \in$ & \multicolumn{2}{|c|}{Ground Truth}\\ \cline{2-3}
    $[\tau_{\cal R}-\delta, \tau_{\cal R}+\delta]$      & Read Speech & Spont\\ \hline
     Read Speech & 8 & 4 \\ \hline
     Spontaneous & 3 & 8 \\ \hline
     \end{tabular}
     \caption{Performance on $23$ audio segments whose ${\cal R} \in [\tau_{\cal R}-\delta, \tau_{\cal R}+\delta]$. 
     %As can be seen 
     $4$ spontaneous speech audio segments were classified as read speech and $3$ read speech segments were classified as spontaneous speech. }
     \label{tab:delta_analysis}
 \end{table}

Very recently, we came across the Archive of L1 and L2 Scripted and Spontaneous Transcripts And Recordings (\allsstar)  corpus \cite{allsstar-db}. We picked up speech data corresponding to $26$ English speakers ($14$ Female and $12$ Male). Each speaker spoke a maximum of $8$ utterances ($4$ spontaneous and $4$ read) in different settings. The $4$ read speech were 
(a) DHR ($20$ formal sentences picked from the Universal Declaration of Human Rights; average duration $106.2$ s) , (b)
HT2 (simple sentences; phonetically balanced which was created for Hearing in Noise Test; average duration $100.5$ s), (c)
LPP ($33$ sentences picked from Le Petit Prince, average duartion $107.1$ s) and (d)
NWS (North Wind and the Sun Passage, average duration $32.8$ s); while the $4$ spontaneous speech utterances were (a)
QNA (Spontaneous speech about anything for $5$ minutes; average duration $317.5$ s), (b)
ST2 (wordless pictures from "Bubble Bubble" used to elicit spontaneous speech; average duration $88.8$ s), (c)
ST3 (wordless pictures from "Just a Pig at Heart"; average duration $78.2$ s), and (d)
ST4 (wordless pictures from "Bear's New Clothes"; average duration $85.2$ s).

\begin{table}[thb!]
    \centering
    \begin{tabular}{|c|c|c|c|c|} \hline
     Gen   &  SpkID & R (DHR, HT2, LPP, NWS)& S (QNA, ST2, ST3, ST4)& (minutes)\\ \hline
     \multirow{14}{*}{F}&   49& 4 (1, 1, 1, 1)&4 (1, 1, 1, 1)& 8 (13.47)\\
     & 51 & 4 (1, 1, 1, 1) & 4 (1, 1, 1, 1) & 8 (16.87)\\
     & 56 & 4 (1, 1, 1, 1) & 4 (1, 1, 1, 1) & 8 (19.29)\\
     & 58 & 4 (1, 1, 1, 1) & 4 (1, 1, 1, 1) &8 (16.73)\\
     & 60 & 4 (1, 1, 1, 1) & 4 (1, 1, 1, 1) &8 (12.32) \\
     & 62 & 4 (1, 1, 1, 1) & 4 (1, 1, 1, 1) &8 (12.78) \\
     & 63 & 4 (1, 1, 1, 1) & 4 (1, 1, 1, 1) &8 (19.42)\\
     & 64 & 4 (1, 1, 1, 1) & 4 (1, 1, 1, 1) &8 (16.06) \\
     & 65 & 4 (1, 1, 1, 1) & 4 (1, 1, 1, 1) &8 (12.70)\\
     & 67 & 4 (1, 1, 1, 1) & 4 (1, 1, 1, 1) &8 (15.04)\\
     & 68 & 4 (1, 1, 1, 1) & 4 (1, 1, 1, 1) &8 (12.91)\\
     & 69 & 4 (1, 1, 1, 1) & 4 (1, 1, 1, 1) &8 (14.90)\\
     & 71 & 4 (1, 1, 1, 1) & 4 (1, 1, 1, 1) &8 (12.87)\\
     & 72 & 4 (1, 1, 1, 1) & 4 (1, 1, 1, 1) &8 (15.67)\\ \hline
     \multirow{12}{*}{M}&   50& 4 (1, 1, 1, 1)&4 (1, 1, 1, 1)&8 (14.28)\\
     & 52 & 4 (1, 1, 1, 1) & 4 (1, 1, 1, 1) &8 (25.4)\\
     & 53 & 4 (1, 1, 1, 1) & 4 (1, 1, 1, 1) &8 (13.27)\\
     & 55 & 4 (1, 1, 1, 1) & 4 (1, 1, 1, 1) &8 (13.27)\\
     & 57 & 4 (1, 1, 1, 1) & 4 (1, 1, 1, 1) &8 (19.26)\\
     & 59 & 4 (1, 1, 1, 1) & 4 (1, 1, 1, 1) &8 (13.60)\\
     & 61 & 4 (1, 1, 1, 1) & 4 (1, 1, 1, 1) &8 (14.37)\\
    % & 71 & 4 (1, 1, 1, 1) & 4 (1, 1, 1, 1) &8 \\
     & 66 & 4 (1, 1, 1, 1) & 4 (1, 1, 1, 1) &8 (14.67)\\
     & 70 & 4 (1, 1, 1, 1) & 4 (1, 1, 1, 1) &8 (12.97)\\
     & 131 & 4 (1, 1, 1, 1) & 2 (1, 1, 0, 0) &6 (11.89)\\
     & 132 & 4 (1, 1, 1, 1) & 2 (1, 1, 0, 0)&6 (12.19)\\
     & 133 & 4 (1, 1, 1, 1) & 2 (1, 1, 0, 0)&6 (12.64)\\ \hline
     Total & 26 (Speakers)& 104 (26, 26, 26, 26) & 98 (26, 26, 23, 23)& 202 (388.9) \\ \hline
    \end{tabular}
    \caption{\allsstar\ corpus details.}
    \label{tab:dataset}
\end{table}

\begin{figure}[thb!]
    \centering
    {\hfill \includegraphics[width=0.45\textwidth]{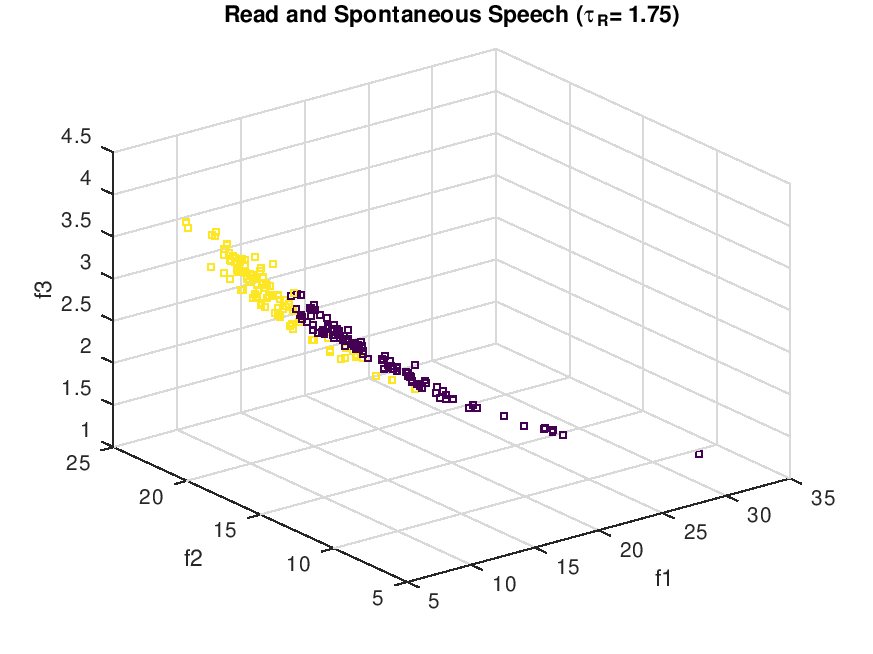}
    \hfill
    \includegraphics[width=0.45\textwidth]{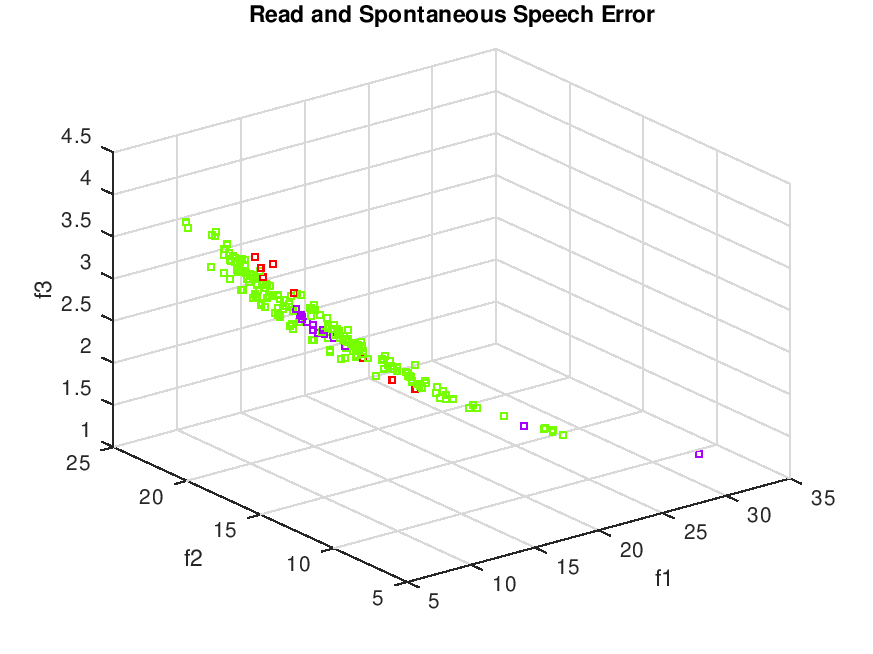}
    \hfill }
    
    {\hfill (a) \hfill (b) \hfill}
    \caption{Classification results on \allsstar. (a) Yellow represents read speech while violet corresponds to spontaneous speech and (b) Green shows the correctly recognized utterances (88.12\%) while red represents read speech recognized as spontaneous and purple shows the utterances corresponding to spontaneous speech which have been recognized as read. }
    \label{fig:allsstar_results}
\end{figure}

In all there were $202$ audio utterances of which $104$ were read utterances and $98$ were spontaneous spoken utterances. Note that in all there should have been $104$ spontaneous utterances; but $2$ spontaneous utterances each were missing from $3$ male participants. Table \ref{tab:dataset} shows the distribution of data from \allsstar. Experiments were carried out on these $202$ audio utterances from $26$ people. We went through the process of passing through audio utterance through the \DS, followed by extraction of three features and computing of ${\cal R}$ as mentioned in (\ref{eq:classification}). The experimental results are shown as a confusion matrix in Table \ref{tab:performance}. As can be observed, the performance of our proposed scheme is $88.12\%$. Figure \ref{fig:allsstar_results} shows the utterances in the feature space $(f_1, f_2, f_3)$ for \allsstar. The classification based on the approach mentioned earlier in this paper is shown in  Fig. \ref{fig:allsstar_results} (a) the utterances classified as read and spontaneous have been marked in yellow and violet respectively.  Figure \ref{fig:allsstar_results} (b) captures the utterances which have been correctly recognised (represented in green). The read utterances mis-recognized as spontaneous is shown in red ($8$ utterances) while the utterances corresponding to spontaneous speech which have been recognized as read have been represented in purple ($16$ utterances).

\begin{table}[thb!]
    \centering
    \begin{tabular}{c|c|c|} \cline{2-3}
          & \multicolumn{2}{c|}{Ground Truth} \\ \cline{2-3}
         & Read & Spontaneous\\ \hline
       \multicolumn{1}{|c|}{Read}  & 88 (84.62\%)& 8 \\ \hline
       \multicolumn{1}{|c|}{Spontaneous} & 16 & 90 (91.84\%)\\ \hline
    \end{tabular}
    \caption{Confusion Matrix. Performance Accuracy on \allsstar\ $88.12\%$.}
    \label{tab:performance}
\end{table}

We analyzed further to understand the mis-recognized utterances. The spontaneous utterances of speakers with ID $49, 56, 58, 60, 71 (2), 57,$ and 
$59$ were mis-recognized as read speech 
%when the utterances were spontaneous and
%recognized as Read (8)
while read utterances with speakers ID $56, 58 (2), 64 (3),$ $69, 71 (2),
 50 (2), 52, 55, 66 (2), 133$ were recognized as being spontaneous.
 %while the utterances were read. 
As shown in Table \ref{tab:error_analysis} we observe that majority of the speakers were mis-recognized either as reading while they had spoken spontaneously (column 1) or as being spontaneous when they had actually read (column 2). Only speakers with SpkID $56, 58$ and $71$ (column 3) were mis-recognized both ways, namely their read speech was recognized as spontaneous and vice-versa. 

% Except for speakers with I the rest of the speakers were either mis-recognized to be spontaneous when they were actually reading or  mis-recognized as reading when they were being spontaneous but not both. 

\begin{table}[thb!]
    \centering
    \begin{tabular}{c|c|c|c|} \cline{2-4}
         %SpkID 
         & Spontaneous $\rightarrow$ Read & Read $\rightarrow$ Spontaneous & Read $\leftrightarrow$ Spontaneous\\ \cline{1-4}
         \multicolumn{1}{|c|}{Female} & $49 (1), 60 (1)$ &  $64 (3), 69 (1)$ & $56 (2), 58 (3), 71 (4)$\\ \hline
         \multicolumn{1}{|c|}{Male} &$57 (1), 59 (1)$ & $50 (2), 52 (1), 55 (1), 66 (2), 133 (1)$& - \\ \hline
    \end{tabular}
    \caption{Mis-recognition based on Speaker ID. The number in parenthesis shows the number of instances.}
    \label{tab:error_analysis}
\end{table}

We observe that the speaker with ID $71$ had ${\cal R} \in [1.63, 1.82]$; we carefully listened to all the utterances and found very less perceptual difference between read and spontaneous utterances. While the read utterances of the speaker with ID $66$ had large silences between sentences (an indication of spontaneous speech) which lead to almost all of the read utterances being recognized as spontaneous.

% Gender	ID	1.752	QNA	ST2	ST3	ST4	
% F	49	R				1	1
% 	56	R		1			1
% 	58	R	1				
% 	60	R			1		1
% 	71	R		1		1	2
% M	57	R	1				1
% 	59	R			1		1
% Total Result			2	2	2	2	8

% read  recognized as Spontaneous

% Gender	ID	1.752	HT2	LPP	NWS	
% F	56	S	1			1
% 	58	S		1	1	2
% 	64	S	1	1	1	3
% 	69	S	1			1
% 	71	S	1	1		2
% M	50	S		1	1	2
% 	52	S		1		1
% 	55	S		1		1
% 	66	S	1	1		2
% 	133	S		1		1
% Total Result			5	8	3	16

%These audio segments were manually annotated by listening to them. 
%We are in the process of scouting for an available data-set specifically build for distinguishing read and spontaneous speech to extensively test our simple, yet {\em novel set of explainable features} to classify speech into read or spontaneous speech. The features selected are very promising because they fit well into the what distinguishes read and spontaneous speech as reported in \cite{Ward_understandingspontaneous}.

\section{Conclusion}
\label{sec:conclusions}
In this paper, we proposed a simple classifier to identify read and spontaneous speech. The novelty of the classifier is in deriving a very small set of features, indirectly from the audio segment. Most of the literature which directly or indirectly address recognition of spontaneous speech have done by analyzing audio signal for determining speech specific properties like intonation, repetition of words, filler words, etc. We derived a small set of explainable features %which are novel because they are based on 
from a string of alphabets derived from the output of the \DS\ speech-to-alphabet recognition engine.  %and not on speech features that are commonly used in speech analysis. 
The features are self explanatory and capture %representative of 
the essential difference between read and spontaneous speech as mentioned in \cite{Ward_understandingspontaneous}.
The derived features are based on {\em how} the utterance was spoken and not on {\em what} was spoken thereby making the features independent of the linguistic content of the utterance.
Experiments conducted on our own data-set (\airdata) and publicly available \allsstar\ shows the classifier to perform very well. 
%The proposed approach 
%shows promise, in spite of the fact that it is simplistic because it is based of a very few {discriminating} features. 
The main advantage of the proposed scheme is that the features are %easily 
explainable and are derived by processing the alphabet string output of $\mbox{\DSm}()$. 
%in addition to being % the features are 
%independent of the performance of the speech recognition engine. %, and they are based on  {\em how} you speak rather than {\em } is spoken. 
It should be noted that while we can categorize our approach as being devoid of deep model training or learning;  the dependency on \deepspeech\ pre-trained deep architecture model (as a black-box) cannot be ignored. 
%Additionally, there is an indirect dependency on this approach working for English language (based on \deepspeech\ pre-trained model used in our analysis). 
% We are in the process of exploring (a) if there is a lower limit of the {\em minimum length of audio} for this analysis to work and (b) if the \deepspeech\ pre-trained model be used for other Indian languages primarily motivated by the fact that the features are independent of the linguistic content of the speech.

\bibliographystyle{splncs04}
\bibliography{references.bib}

\end{document}